\documentclass[]{interact}
\usepackage{graphicx}
\usepackage[T1]{fontenc}
\usepackage{amssymb}
\usepackage{amsthm}
\usepackage[official]{eurosym}
\usepackage{booktabs}
\usepackage{pifont}
\newcommand{\cmark}{\ding{51}}%
\newcommand{\xmark}{\ding{55}}%

\usepackage{natbib}
\bibpunct[, ]{(}{)}{;}{a}{}{,}

\theoremstyle{plain}

\theoremstyle{definition}

\theoremstyle{remark}

\begin{document}


\title{End-to-End Quality of Service Management in the Physical Internet: a Digital Internet Inspired Multi-Domain Approach}

\author{
\name{F. Phillipson
}
\affil{TNO, The Netherlands}
}

\maketitle

\begin{abstract}
For the layer ‘System Level Functionality’ of the Phyisical Internet, it is needed to estimate end-to-end performance characteristics of transportations that visit multiple logistic domains. This paper proposes an approach based on a Digital Internet functionality: a combination of a Service Level Agreement registry and a Quality of Service processor. Existing SLA-calculus gives tools for the QoS-processor  to  combine  the  SLA-parameters  for  all  possible  end-to-end paths and gives the QoS-processor the possibility to propose the best path given the required performance.  A realistic implementation is proposed using a multi objective/constraint approach and a related communication form between the domain owner and the QoS Processor.
\end{abstract}


\section{Introduction}
\label{S:1}
\begin{figure}
    \centering
    \includegraphics[width=14.5cm]{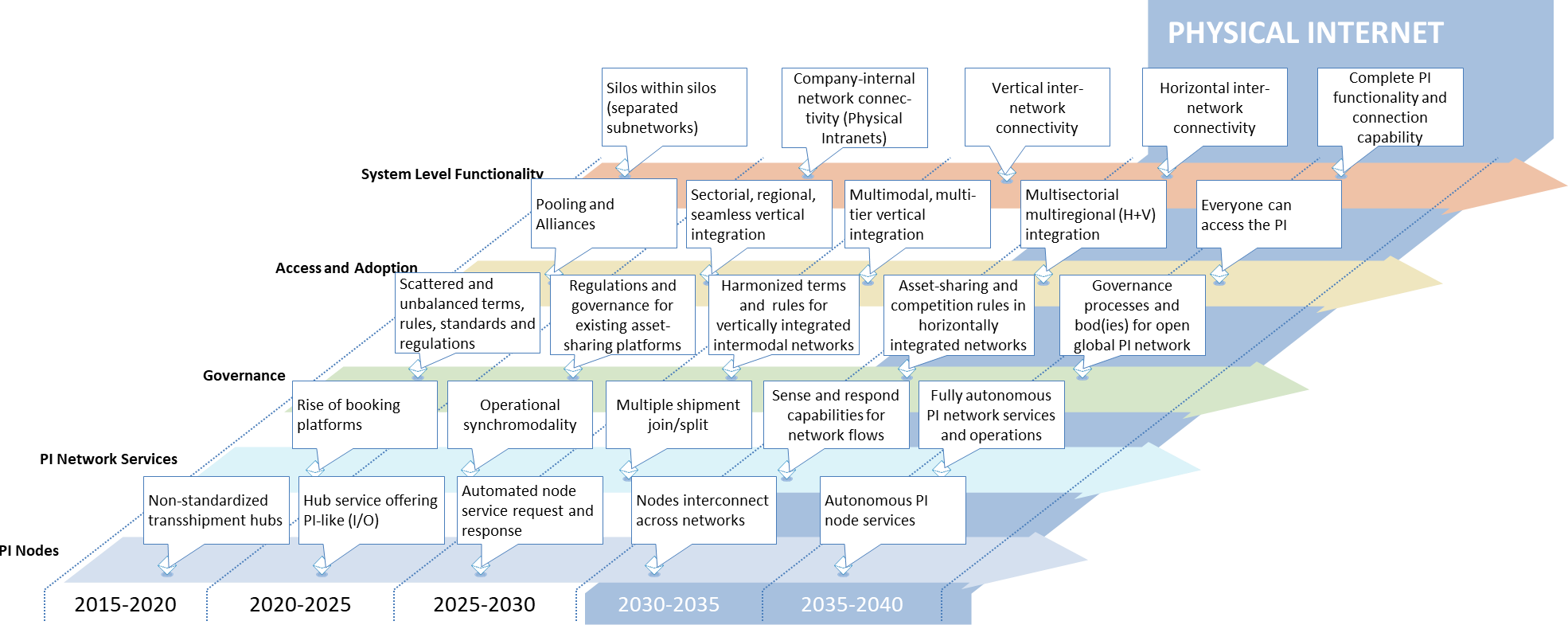}
    \caption{The Physical Internet roadmap from \cite{alice}.}
    \label{fig:alice}
\end{figure}

The Physical Internet (PI) is a logistic concept that aims at realising full interconnectivity (information, physical and financial flows) of several (private) freight transport and logistics services networks and make them ready to be seamlessly used as one large logistics network (\cite{alice}). The concept of the Physical Internet was introduced by \cite{montreuil2011toward}. The goal is to create an open global logistics system founded on physical, digital and operational interconnectivity through encapsulation, interfaces and protocols (\cite{montreuil2013physical}) to obtain a more efficient logistic system in terms of costs (\cite{venkatadri2016physical}), carbon emissions and carbon-free city logistics (\cite{cipres2019physical,ballot2018improving}). A lot of work has been done on this topic since then, of which an overview can be found in the literature overviews of \cite{treiblmaier2016conceptualizing}, \cite{sternberg2017physical} and \cite{treiblmaier2020physical}. The PI does not directly manage physical goods but rather manages the shipping of containers that store goods, just as Digital Internet (DI) uses packets to store user data. For the PI to become full-fledged, numerous elements need to be coordinated, including physical objects, such as PI modular containers and PI transit centres, and more abstract concepts, such as legislation and business models (\cite{treiblmaier2020physical}). The Alice/Sense `Roadmap to Physical Internet' in the \cite{alice} explains and describes the development of PI in the next years, using five specific areas: 
\begin{enumerate}
\item PI Nodes - Roles of physical nodes and operational model for physical nodes; 
\item PI Network Services – PI protocol stack and network management; 
\item Governance - Governance concepts, bodies, regulations and trust building measures; 
\item Access and Adoption – Benefits of PI and mental shift towards PI;
\item System Level Functionality - PI architecture, building blocks and information exchange.
\end{enumerate}
Each of the areas in the roadmap has starting in 2015-2020 and shows the possible developments in `generations' until 2040, see Figure \ref{fig:alice}. Generations define possible evolutions towards PI and can be scenarios or parts of PI-like implementations. PI-like operation will be established around 2030, the developments from 2030 to 2040 focus on improvements on the way to fully autonomous PI operation. 

In this paper we will look at the highest level, the System Level Functionality. In the \cite{alice} is stated that the basic operational premise of the Physical Internet is similar to the one that has served the Digital Internet so well over its almost 50 years of operations and growth. That premise is that simplicity and openness must drive all system level protocols and controls. Such a premise provides users with the opportunity to innovate without burdening them with overhead that inhibits creativity. With respect to the PI, this premise implies that the PI needs only a small amount of information concerning the shipments that are entrusted to it for delivery to ensure their delivery. The protocols to be employed by the PI also should reflect the minimalist nature of the protocols employed in the DI.  While the DI’s protocols have grown more complex over the years, the PI should attempt to start out as the DI did with an extremely simple protocol structure and allow circumstances to drive the modification of these originating control processes.  As a starting point, the protocols for the PI should control how nodes forward shipments onwards, how costs are collected, how errors and exceptions are to be handled, how end-to-end control is to be maintained, how node-to-node control is to be maintained, how congestion is to be managed, how differing physical layers are to be managed, how QoS parameters should be handled, and how users can access data concerning each of these managed activities. The Alice roadmap in the \cite{alice} gives an example of how the PI could work:
\begin{enumerate}
    \item A shipper is interested in sending a consignment of goods from Barcelona to Hamburg.  The shipper contacts their freight forwarder and asks the freight forwarder what the cost will be for the shipment providing the forwarder with delivery information, shipment information and required time of delivery. 
    \item The freight forwarder queries the PI to obtain an estimate of costs for the shipment, given the shipments characteristics and delivery time.
    \item The PI examines the end-to-end route that the goods most likely will travel (based on historical information concerning origin-destination routings), integrates existing loads anticipated for the dates in question over the routes considered, incorporates node costs, transport costs, and any other pertinent costs for handling along the most likely route, and any known maintenance information to arrive at a preliminary cost estimate for the shipment. 
\end{enumerate}

We will now give a realistic implementation of step 2 and 3, inspired by an approach proposed for the DI. In \cite{blom2011system} a Service Level Agreement (SLA) registry is proposed which provides Quality-of-Service (QoS) data. This data is handled by a QoS processor, based on SLA-calculus as proposed in \cite{kooij2006sla}, \cite{van2005user} end \cite{van2006modelling}, resulting in predictions of end-to-end QoS parameters used for selection and/or ranking actions for new session requests.
In the next section this original DI approach is explained in detail. Next, in Section \ref{S:3} the proposed concept in the PI environment is elaborated on. An example implementation of this concept is shown in Section \ref{S:4}. The paper ends with some conclusions.

\section{End-to-End QoS in multi-domain Telecommunication networks}
In telecommunication multi-domain end-to-end QoS was a hot topic already in 2002 (\cite{kamienski2003strategies}). Each domain is a telecommunication network of a specific Service Provider (SP), each with its own characteristics, QoS-mechanism and service levels. For a specific service (for example Voice-over-IP) a connection has to be established over multiple domains between the origin and the destination of the call. The service provider of the originating domain wants some guarantee about the end-to-end QoS, as he promises some service level to his clients. 

In \cite{kooij2006sla} is stated that an effective and increasingly popular means to deal with QoS in multi-domain environments is to negotiate Service Level Agreements (SLAs) between the different domain owners. In this context, a key question to be addressed by a SP is “What combination of SLAs should be agreed upon by the SP and the respective network domain owners to achieve a certain predefined end-to-end QoS level?”. Note that an SP would not subscribe to a costly premium service class in one of the intermediate backbone networks, when there appears to be an inevitable bottleneck with respect to the achievable end-to-end QoS in, e.g., the wireless access. An example of interdomain dynamic service negotiation, which is an important step that must be considered prior to the physical resource provisioning, can be found in \cite{kamienski2003strategies}. In more general terms, the service provider should try to distribute the involved ‘performance budgets’ (e.g., packet loss, delay) over the underlying domains, such that the envisioned end-to-end QoS level can be realised against minimal costs. It is clear that this ‘cost optimisation’ problem requires the ability to determine end-to-end QoS guarantees from the performance parameters specified in the SLAs. This mapping of a set of per-domain SLAs to an associated (achievable) end-to-end QoS level is called SLA calculus, examples of which can be found in \cite{van2005user}, \cite{kooij2006sla}, \cite{van2006modelling} and \cite{vastag2014sla}. These papers show for various services (voice-over-IP, web serving) what key performance indices (KPIs) per domain should be shared and how these can be translated to the end-to-end performance and in some cases to some perceived quality measure. This perceived quality measure is a prediction of how more technical parameters such as delay, response time etc., are perceived by the end-user and translated to a Mean Opinion Score (MOS), as in \cite{yamamoto1997impact}. Additional, in the operational setting, call admission control can be used to guarantee the SLAs, as proposed in the framework of \cite{burakowski2008provision}. They present a framework to provision end-to-end QoS in heterogeneous multi-domain networks that was implemented in EuQoS system and tested in Pan-European research network.

If we know what information has to be shared to be able to calculate the end-to-end performance, the following questions arise: where this information is stored? Who can access this information/data to perform the calculations? For both questions the answer can be `each domain owner individually', what we will call a distributed approach, and `some orchestrator' what will be called a global approach. This leads to four main areas:
\begin{enumerate}
    \item Distributed/Distributed: if both data and the calculation are distributed, for each call all possible used domains have to be contacted, leading to a huge overhead on communication.
    \item Distributed/Global: if data is distributed and the calculation is done globally, some orchestrator has to take this role and communicate with all domains each time a request is made. 
    \item Global/Distributed: if data is global and the calculation is distributed, the domain owner can request the database and perform the calculation.
    \item Global/Global: now all data is centralised and the domain owner gets a single answer to his request.
\end{enumerate}
When data is shared, the question about confidentiality arises: do we want to share this data. Also, when there is a single orchestrator, it has to be trusted by all parties, both in data confidentiality as well in giving the best, unbiased, answer to the request.

In \cite{blom2011system} a Global/Global system is described. Preconditions that are set in this work for a such system are:
\begin{enumerate}
    \item Domain owners should be able to make their own decisions;
    \item No new protocol should be needed for such a system;
    \item The solution should be scalable;
    \item Commercial Service Level Agreement (SLA) information must be kept confidential.
\end{enumerate}

This system proposes a combination of a QoS processor and an SLA registry, predicting the end-to-end QoS value of the paths between the originating and destination location, based on the SLA information as stored in the SLA registry. The system ranks the relevant paths end-to-end QoS value order and selects the path(s) having the best end-to-end QoS prediction. The SLA information is kept confidential as the SLA registry does not exchange the SLA information itself, only the ranking and the information about the requested end-to-end QoS will be obtained. This information is given using a MOS value. The main reason that the system generates an ordered list instead of just prescribing the 'next to go' domain is that the QoS processor only calculated long-term predictions without taking into account the dynamics of the current status of each individual domain. This lack of dynamics is off course depending on the time interval of renewing the SLA information within the registry. The method can be implemented as a distributed database and processor, comparable to DNS servers in the DI. As indicated, this method promises confidentiality, but asks trust of the central orchestrator. As alternative, a decentral method based on secure multiparty computation as presented in \cite{cramer2015secure} can be used. In that case, a method is used for parties to jointly compute a function over their inputs, while keeping those inputs private.

\section{End-to-End QoS in multi-domain Physical Internet} \label{S:3}
Comparable to the approach in \cite{blom2011system} we propose a Global/Global system for the Physical Internet. Now, a domain is the logistic network of a single Logistic Service Provider (LSP) who offers transportation services within its network and as part of a multi LSP domain delivery. The main conditions we assume for the QoS system are:
\begin{enumerate}
    \item Domain owners should be able to make their own decisions; 
    \item The solution should be scalable; 
    \item Commercial Service Level Agreement (SLA) information must be kept confidential. 
\end{enumerate}
For this we propose an SLA database or registry in which domain owners upload their SLA parameters and a QoS processor which processes the QoS related parameters from the SLA registry for predicting an end-to-end QoS value that is representative for the path between the originating location and the destination location. In many cases there will be (a plurality of) alternative paths between the originating location and the destination location. Then the QoS processor is adapted for predicting a plurality of end-to-end QoS values. The QoS processor calculates (predicts) the end-to-end QoS value of the paths between the originating and destination location, based on the SLA information as stored in the (common) SLA registry and ranks the relevant paths end-to-end QoS value order or, alternatively, selects the path having the best end-to-end QoS prediction. This action can be done on request, meaning each time an LSP demands an end-to-end QoS value. However, as \cite{blom2011system} also proposes, an alternative approach can be calculating and storing QoS values for all or a selection of paths in advance or store path values in the SLA registry when calculated on request. This leads to a situation where for a specific request the information can be retrieved from the database directly. Re-computation of the end-to end QoS predictions for some paths may then be needed only when a domain's SLA changes. 

The implementation of the QoS processor depends on the chosen SLA parameters. One could think of average, minimal and maximal delivery times, order accuracy\footnote{Order accuracy measures the amount of orders that are processed, shipped and delivered without any incidents on its way.}, number of modality changes, cost, $CO_2$ emissions, delivery on time percentage etc. How the convolution of the SLA parameters is done over multiple domains also depends on the specific parameter and the assumed underlying (mathematical) model. In some cases simple addition and subtraction is sufficient or sums of stochastic variables can be calculated easily in the case of assumed underlying (log) normal or Poisson distributions. In other cases more sophisticated techniques are required, when more complex probability distributions are assumed, for example in the form of queuing systems with batch arrivals, modelling the handling by ships or trains within a domain. Techniques from \cite{vastag2014sla} can be of inspiration here. The QoS processor then calculates the convolution value of all the $n$ SLA parameters per end-to-end path resulting in the vector $( x_1,...,x_n)$, where $x_i$ is the resulting end-to-end SLA parameter for the $i$-th SLA parameter.

In the DI case, there exists a single value, MOS, that depicts the perceived quality determined by multiple underlying, physical SLA parameters. In logistics, there does not exist such a single (perceived) quality parameter. Therefor, a multi objective/constraint approach is proposed, where the domain owner can select and define:
\begin{enumerate}
    \item SLA parameters that have to be of a certain minimum/maximum value, the constraints, and
    \item SLA parameters that have to be optimised, asking for coefficients to weight their perceived relative importance, the objectives.
\end{enumerate}

If we have $n$ SLA parameters, we propose to define the input for the QoS processor in the form:
$$ C=(C_1,...,C_n) $$
where $C_i$ is the QoS processor command for SLA parameter $i$. This command can be `$=c_i$', `$>c_i$', `$\geq c_i$', `$<c_i$' or `$\leq c_i$' in the case of a constraint parameter and `$w=w_i$' in the case of a objective parameter. If we separate all SLA parameters into the sets $I_o$ for the objective parameters and $I_{c=}$, $I_{c<}$, $I_{c\leq}$, $I_{c>}$ and $I_{c\geq}$, then the formal optimisation problem based on $C$ equals:
$$ \max \sum_{i \in I_o}  w_i x_i, $$
under the constraints
\begin{align*}
x_i=c_i, \quad \forall i \in I_{c=},\\
x_i<c_i, \quad \forall i \in I_{c<}, \\
x_i\leq c_i, \quad \forall i \in I_{c\leq}, \\
x_i>c_i, \quad \forall i \in I_{c>}, \\
x_i\geq c_i, \quad \forall i \in I_{c\geq}. 
\end{align*}

\noindent As example, the command
$$ C=(=10, >20, w=5, w=1) $$
will result in the optimisation problem
$$ \max 5 x_3 + x_4, $$
under the constraints
\begin{align*}
x_1=10, \\
x_2>20.
\end{align*}

This proposed methodology meets the conditions that we requested in the beginning of this section. The ranking and the composed results make sure that the first and last condition are met: domain owners can make their own decisions and the SLA information is kept confidential. The scalability can be realised by pre-processing and storing paths, reducing the update frequency of SLA parameters and distributed implementation of the SLA registry and QoS processor, in the same way this is done for the DI. To make this work, it should be integrated in a Internet-of-Things service-oriented architecture environment, as proposed in \cite{tran2020towards}, \cite{kim2019information}, or \cite{acciaro2020technological}.

\begin{figure}
    \centering
    \includegraphics[width=13.5cm]{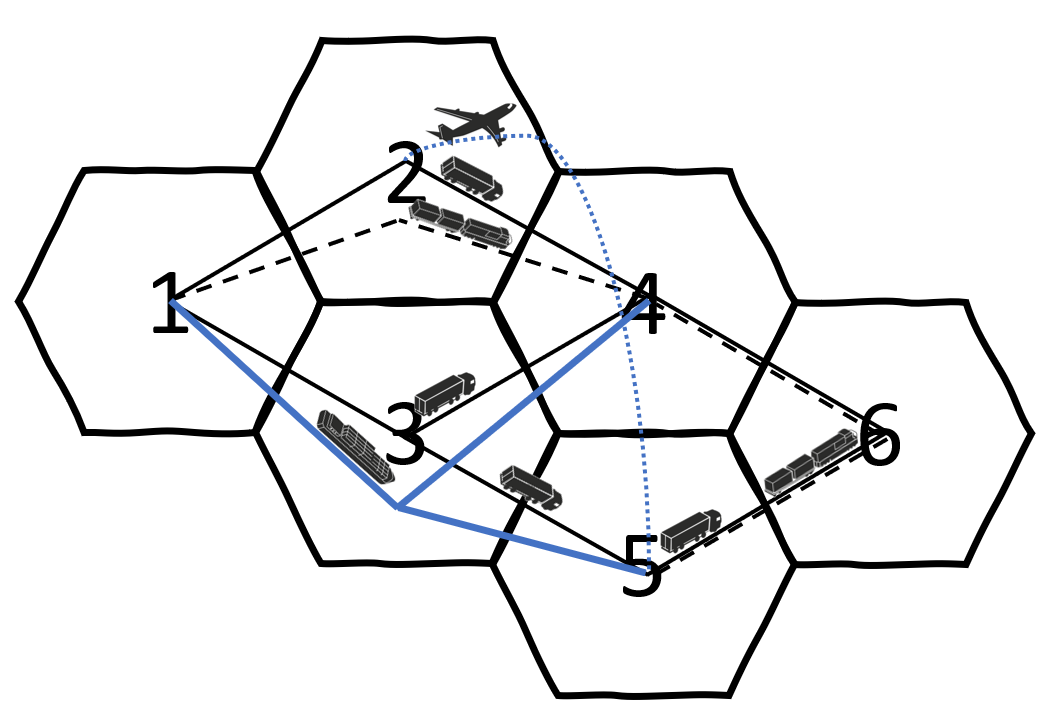}
    \caption{Example logistic network with 6 domains.}
    \label{fig:0}
\end{figure}

\begin{figure}
    \centering
    \includegraphics[width=9cm]{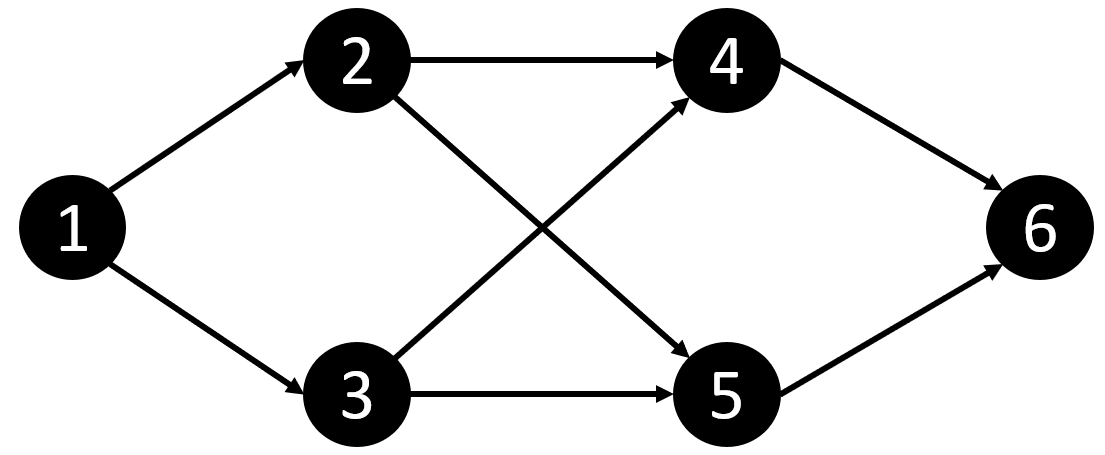}
    \caption{Schematic logistic network with 6 domains.}
    \label{fig:1}
\end{figure}

\section{Example of a Multi Objective QoS Processor} \label{S:4}
Looking at the operations in logistics, examples of important SLA parameters are cost, on time delivery and $CO_2$ emissions. This means that in practice a combination of these SLA parameters is used for steering decisions and indeed a multi objective problem is defined, where the solutions give a score on all the SLA parameters and their total score is a weighted combination. We assume a network of domains as depicted in Figure \ref{fig:0} and schematic in Figure \ref{fig:1}. Note that domains do not require to be geographically separated, a domain stands for the logistic network of a specific LSP. The owner of domain 1 wants to ship a container to domain 6. Each domain owner should register in the SLA registry for each pair of neighbouring domains:
\begin{itemize}
    \item the cost for shipping a container via its domain between the neighbouring domains;
    \item the  $CO_2$ emissions for shipping a container via its domain between the neighbouring domains;
    \item the average and variance transportation time, assuming normal distributed transportation times per domain.
\end{itemize}

In the case of a multi objective environment, it will not suffice for domain owner to report a single SLA value for each pair of neighbouring domains. Instead, a number of Pareto efficient solutions per pair should be determined. A Pareto efficient (or Pareto optimal) solution is a solution for a multi-objective problem where there is no other solution with a better score for one of the objectives, without a worse score for one of the other objectives. If we assume that cost and $CO_2$ emissions have positive correlation, meaning that a more expensive solution, e.g., trucking, has a higher $CO_2$ emission than a less expensive solution, e.g., transport by barge, and that they both have a negative correlation with transportation time, it suffices to report the solution with the lowest cost and the solution with the lowest transportation time per domain. An example is found in Table \ref{tab:1}. Here each intermediate node in the network of Figure \ref{fig:1} reports two Pareto efficient solutions per connection.

\begin{table}[]
    \centering
    \begin{tabular}{cccccc} \toprule
   &&&& \multicolumn{2}{c}{Transportation Time (TT)}\\
Path & Objective    &	Costs (\euro{}) &	$CO_2$ (kg) &	Average (h) & 	Variance\\ \hline
2$\to$4  & TT             &	100&	100&	24&	4\\
	&  Costs             & 80&	60&	35&	12\\
2$\to$5& TT	            &90&	90&	20&	4\\
	& Costs              &80&	70&	36&	16\\
3$\to$4& TT	            &120&	110&	26&	4\\
	& Costs              &100	&50&	50&	15\\
3$\to$5& TT	            &70	&90&	16&	6\\
	& Costs              &90&	70&	36&	16\\
4$\to$6& TT	            &75&	60&	20&	10\\
	& Costs              &70&	50&	25&	15\\
5$\to$6& TT	            &80&	55&	18&	5\\
	& Costs              &75&	40&	22&	10    \\ \toprule
 
    \end{tabular}
    \caption{Items of SLA register; multiple lines per pair of neighbouring domains.}
    \label{tab:1}
\end{table}

Based on this input, the QoS Processor is able to calculate the total costs, $CO_2$ emission and the probability density of the transportation time. Assume that the owner of domain 1 communicates to the QoS processor that he wants to ship a container to domain 6, having preference weights for cost and emission of $w_1=\frac{3}{5}$ and $w_2=\frac{2}{5}$ respectively, and a transportation time of less than 60 hours with 60\% probability,
$$(w_1=\frac{3}{5}, w_2=\frac{2}{5}, >60\% ), $$
the QoS could rank the paths as depicted in Table \ref{tab:2}. Here one could consider to translate the weighted cost into a value that cannot be traced back to the original SLA parameter values.

\begin{table}[]
    \centering
    \begin{tabular}{ccccccc} \toprule
Rank&	Weighted cost&	Path&	Cost&	Emission&	Probability & \\ \hline
1&	134&	2 4 6&	150&	110&	50\% &\xmark\\
2&	137&	2 5 6&	155&	110&	65\% &\cmark\\
3&	139&	3 5 6&	145&	130&	100\% &\cmark\\
4&	141&	2 4 6&	155&	120&	86\% &\cmark\\
5&	142&	3 4 6&	170&	100&	0\% &\xmark\\
6&	143&	3 5 6&	165&	110&	65\% &\cmark\\
7&	146&	2 5 6&	160&	125&	90\% &\cmark\\
8&	148&	3 5 6&	150&	145&	100\% &\cmark\\
9&	149&	3 4 6&	175&	110&	2\% &\xmark\\
10&	151&	2 5 6&	165&	130&	100\% &\cmark\\
11&	152&	3 5 6&	170&	125&	90\% &\cmark\\
12&	160&	2 5 6&	170&	145&	100\% &\cmark\\
13&	162&	2 4 6&	170&	150&	99\% &\cmark\\
14&	169&	2 4 6&	175&	160&	100\% &\cmark\\
15&	178&	3 4 6&	190&	160&	98\% &\cmark\\
16&	185&	3 4 6&	195&	170&	100\% &\cmark\\ \toprule
    \end{tabular}
    \caption{Resulting ranking of QoS Processor and indication of allowed solutions.}
    \label{tab:2}
\end{table}

The solution with the lowest cost is not allowed. This solution does not meet the transportation time probability. The best, allowed, solutions can now be communicated to the owner of domain 1. 

\section{Conclusions}
The Physical Internet concept needs to be completed in the next decades, following the `Roadmap to Physical Internet' \cite{alice}. For the layer `System Level Functionality', the PI needs to estimate end-to-end performance characteristics of transportations that visit multiple LSP domains. Where Physical Internet is inspired by the Digital Internet, also for this function an DI inspired approach is chosen: a combination of a SLA registry and a QoS processor, both on the global level of the PI. SLA-calculus gives tools for the QoS-processor to combine the SLA-parameters for all possible end-to-end paths and gives the QoS-processor the possibility to propose the best path given the required performance. A realistic implementation is proposed using a multi objective/constraint approach and a related communication form between the domain owner and the QoS Processor. 

\bibliographystyle{tfcad}
\bibliography{sample.bib}
\end{document}